\documentclass[%
superscriptaddress,
 frontmatterverbose, 
 reprint,
 amsmath,amssymb,
 prl,
showkeys,
]{revtex4-2}

\usepackage{graphicx}
\usepackage{dcolumn}
\usepackage{bm}
\usepackage{mathptmx}
\usepackage{ulem}
\usepackage{mathtools}
\usepackage{subcaption}
\usepackage{hyperref}
\usepackage{psfrag}
\usepackage{graphicx}
\usepackage[dvipsnames]{xcolor}
\definecolor{inkscape_green}{RGB}{19, 158, 41}

\begin{document}

\preprint{APS/123-QED}

\title{Willis coupling-induced acoustic radiation force and torque reversal}


\author{Shahrokh Sepehrirahnama}
\email{shahrokh.sepehrirhanama@uts.edu.au} 
\affiliation{%
 Centre for Audio, Acoustics and Vibration, University of Technology Sydney, Sydney, Australia
}

\author{Sebastian Oberst}
\email{sebastian.oberst@uts.edu.au} 
\altaffiliation[Also at ]{School of Engineering and Information Technology, University of New South Wales, Canberra, Australia}
\affiliation{%
Centre for Audio, Acoustics and Vibration, University of Technology Sydney, Sydney, Australia
}%

\author{Yan Kei Chiang}%
\altaffiliation[Also at ]{Centre for Audio, Acoustics and Vibration, University of Technology Sydney, Sydney, Australia}
\affiliation{%
 School of Engineering and Information Technology, University of New South Wales, Canberra, Australia
}%

\author{David A.~Powell}
\affiliation{%
 School of Engineering and Information Technology, University of New South Wales, Canberra, Australia
}%


\date{\today}

\begin{abstract}
Acoustic meta-atoms serve as the building blocks of metamaterials, with linear properties designed to achieve functions such as beam steering, cloaking and focusing.
They have also been used to shape the characteristics of incident acoustic fields, which led to the manipulation of acoustic radiation force and torque for development of acoustic tweezers with improved spatial resolution. 
However, acoustic radiation force and torque also depend on the shape of the object, which strongly affects its scattering properties.
We show that by designing linear properties of an object using metamaterial concepts, the nonlinear acoustic effects of  radiation force and torque can be controlled.
Trapped objects are typically small compared to the wavelength, and are described as particles, inducing  monopole and dipole scattering.
We extend such models to a polarizability tensor including Willis coupling terms, as a measure of asymmetry, capturing the significance of  geometrical features. 
We apply our model to a three-dimensional, sub-wavelength meta-atom with maximal Willis coupling, demonstrating that the force and the torque can be reversed relative to an equivalent symmetrical particle.
By considering shape asymmetry in the acoustic radiation force and torque,  Gorkov's fundamental theory of acoustophoresis is thereby extended.
Asymmetrical shapes influence the acoustic fields by shifting the stable trapping location, highlighting a potential for tunable, shape-dependent particle sorting.

\end{abstract}

\maketitle

\begin{figure}[t]
    \centering
    \psfrag{La}[l][l][1]{(a)}
    \psfrag{a}[t][c][1]{$a$}
    \psfrag{W}[l][l][1]{\textcolor{white}{$W$}}
    \psfrag{Lb}[l][l][1]{(b)}
    \psfrag{PN}[l][l][1]{\textcolor{blue}{P.N.}}
    \psfrag{VN}[l][l][1]{V.N.}
    \psfrag{Force}[r][r][0.9]{Force}
    \psfrag{T}[c][c][0.9]{\textcolor{inkscape_green}{Torque}}
    \psfrag{ST1}[l][l][1]{\textcolor{blue}{Stable trap}}
    \psfrag{ST2}[l][l][1]{\textcolor{red}{Stable trap}}
    \includegraphics[width=\columnwidth]{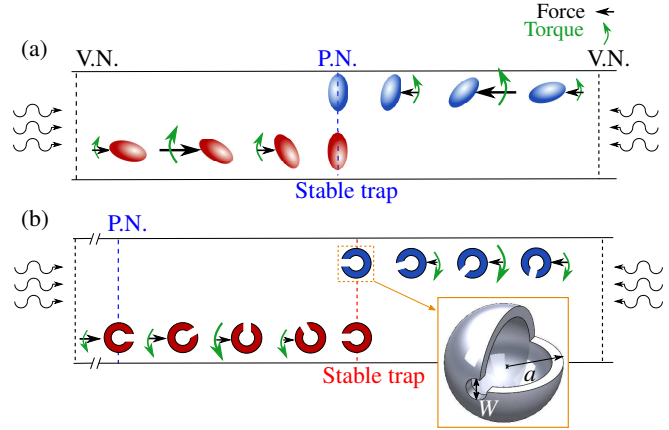}
    \caption{Acoustic trapping of objects with (a) zero and (b) non-zero Willis Coupling, corresponding to symmetric and asymmetric shapes, respectively.
    The asymmetry in the shape affects the locations of an acoustic trap and the stable orientation of the objects in it.
    A meta-atom based on a Helmholtz resonator exhibiting such non-conventional acoustic trapping.
    P.N. and V.N. stand for Pressure Node and Velocity Node of an incident acoustic field.
    Stable trap refers to the location of zero radiation force with negative force gradient.}
   \label{fig:schematic_sph_MA}
\end{figure}

Acoustic radiation force and radiation torque are the physical quantities underlying acoustophoresis - the manipulation of sub-wavelength objects by an incident acoustic field \cite{af_bruus1, af_bruus7, af_Laurell, af_bruus10, af_dual2012, af_wiklund, Lim_2011, Lim_2018}.
Applications such as ultrasonic sorting, separation and levitation have been developed by applying force and torque to objects such as biological cells \cite{Lim_2011, Laurell2012, Hill_2014, antfolk2015_CTC, DW2015, Lim_2016, DW2019, memoli2020AcLevMeta}.
Considering their magnitude relative to gravitational force and fluid drag, acoustic radiation force and torque are most suitable for practical manipulation of sub-wavelength objects, with sizes ranging from sub-micrometers to a few millimeters \cite{King_34, Yosioka_55, Gorkov_62, Doinikov1994_JFM, Bruus_2012, silva_Bruus, marston2006axial, mitri2015, garcia2014experimental, Lim_2018}.
Existing theoretical derivations of the acoustic radiation force and radiation torque use simple shapes such as spheres, spheroids, cylinders and disks, leading to simple expressions which show the influence of the object's volume, material and aspect ratio \cite{Doinikov1994_Proc,foresti2012ellSpheroid,FBW2015spheroid,mitri2015ellCyl,wei2004cyl,xie2004ARFdisc,garbin2015ARFdisk}.
\textcolor{black}{
This simplifying assumption neglects asymmetry, which is a potential degree of freedom to tune acoustophoretic processes, for example  manipulation of a heterogeneous mixture of sub-wavelength objects, as illustrated in Fig.~\ref{fig:schematic_sph_MA}.} 
\textcolor{black}{
Neglecting shape asymmetry, Gorkov's theory of acoustophoresis shows that small objects can be trapped at pressure or velocity nodes, cf. Fig.~\ref{fig:schematic_sph_MA}(a), depending on their dominant scattering mode \cite{Bruus_2012, shahrokh2020PRE}.} 
However, when shape asymmetry is accounted for, the trapping locations may shift away from the pressure node, as shown in Fig.~\ref{fig:schematic_sph_MA}(b).

Considering the key role of acoustic radiation force and torque in the ultrasonic range, it is important to capture the effects of geometric asymmetry on these force and torque fields.
Recent advances in asymmetric metamaterials indicate growing interest in designs based on engineered geometries, either as a material of a single meta-atom for beam steering \cite{Alu2018maxWC, jordaan2018, anton2019, YK2020arraye} or used as a material made of an array of meta-gratings for beam splitting \cite{ni2019metagratings}.
Such asymmetric shapes are characterised by Willis coupling, which describes the monopole and dipole moments induced by the incident velocity and pressure fields, respectively \cite{Alu2017WCorigin}.
This coupling has a theoretical upper limit and approaches zero as the object approaches a mirror-symmetric configuration \cite{Alu2018maxWC}.
By geometrical modifications such as adding protruding holes and internal cavities, the Willis coupling and its effect on the acoustic field can be controlled \cite{Alu2017WCorigin, norris2018, anton2020, YK2020arraye}. 
In particular, a two-dimensional C-shaped Helmholtz resonator can achieve the maximum Willis coupling near its resonance, or other arbitrary values by having multiple apertures of different sizes \cite{anton2019}.

To investigate the influence of Willis coupling on the acoustic radiation force and torque, we consider a three-dimensional version of the Helmholtz resonator based meta-atom, as illustrated in Fig.~\ref{fig:schematic_sph_MA}(b). 
\textcolor{black}{The cavity can be considered as a secondary scatterer that is employed to tune the overall scattering response of the meta-atom.} 
\textcolor{black}{
Since our design targets the ultrasonic range, narrow and long passages are avoided due to their role in thermo-viscous losses \cite{YK2020arraye}.
The losses for airborne acoustic meta-materials with narrow passages depend on the depth of the viscous boundary layer, which is inversely proportional to the square root of the frequency and is of the order of micrometers \cite{zhang2017_tv}.
Here, thermo-viscous losses are neglected for the theoretical investigation of shape-controlled meta-material behavior.
This assumption is validated in the Supplementary Notes \cite{supp_notes}.} 

The 3D meta-atom has an internal cavity, occupying 60$\%$ of the volume, and one aperture of width $W=0.2a$ with $a$ being the outer radius of the spherical shell, which acts as the normalization parameter, cf. Fig.~\ref{fig:schematic_sph_MA}(b).
These parameters are tuned to achieve maximum Willis coupling along a single axis \cite{anton2019}.
The values of $a$ are varied from $0.01\lambda$ to $0.16\lambda$\textcolor{black}{, corresponding to $ka\approx 0.06$ to $0.96$, respectively, to investigate the effects of shape asymmetry in the Rayleight limit of $ka\ll 1$.
Beyond $ka>1$, higher order terms, e.g. quadrupole, octopole etc., are required for a more accurate representation of scattered pressure and the formulation of acoustic radiation force and torque.} 
%
%
We assume that the meta-atom behaves as a sound-hard and immovable object in the acoustic domain.
This allows comparison with existing radiation force theory, which predicts, for a plane standing wave, sound-hard immovable spheres or spheroids are always pushed towards the pressure nodes. 
These are the stable acoustic trap locations with zero force and negative force gradient, as \textcolor{black}{illustrated} in Fig.~\ref{fig:schematic_sph_MA}.
\textcolor{black}{For \textcolor{black}{studying} radiation torque, a sound-hard immovable prolate spheroid of sub-wavelength size is used as a reference  \cite{leao2021_cyl_spheroid, leao2020_spheroid}}.
\textcolor{black}{
The radiation force and torque applied to axisymmetric shapes were studied analytically and numerically \cite{marston2017_phase_shift, lima2020_spheroid, jerome2021_spheroid, lopes2020_spin, Silva_2021_RBC} in different beam types \cite{zhang2011_bessel_neg, zhang2011_angular, zhang2018_reversals, zhang_2019_bessel_trap, zhang2021_phase_bessel}; however, the locations of acoustic traps were the same as in Gorkov's theory.
Here, we demonstrate the correlation between shape asymmetry and the location of an acoustic trap due to the force reversal phenomenon.} 


The incident field is taken as a plane standing wave with $p_i = P_a\cos(kz)e^{-j\omega t}$, and the meta-atom is represented by its monopole-dipole polarizability tensor $\pmb{\alpha} = \begin{bmatrix} \alpha_{pp} & \pmb{\alpha}_{pv}^T\\
\pmb{\alpha}_{vp}& \pmb{\alpha}_{vv}\end{bmatrix}$. The acoustic radiation force and torque are derived as the sum of  symmetric and asymmetric terms \cite{shahrokh2021_WC_ARF}, $\mathbf{F}=\mathbf{F}_{\text{sym}}+\mathbf{F}_{\text{asym}}$ and $\mathbf{T} = \mathbf{T}_{\text{sym}} + \mathbf{T}_{\text{asym}}$ with
\begin{align}
    \begin{split}
        \mathbf{F}_{\text{sym}} =& \frac{kE_i}{\rho_f}\Big(\frac{\Re{\big[\alpha_{pp}\big]}}{\kappa_f}\mathbf{e}_z - k c_f\Im{\big[\pmb{\alpha}_{vv}\mathbf{e}_z\big]}\cdot \mathbf{e}_z\mathbf{e}_z\Big) \sin(2kz),\\
        \mathbf{F}_{\text{asym}} =& \frac{kE_i}{\rho_f} c_f\Im{\big[2\pmb{\alpha}_{pv}\cdot\big(\mathbf{e}_z\mathbf{e}_z\big)\big]}\cos(2kz),
    \end{split}
    \label{eq:force_exp_PSW}
\end{align}
\begin{align}
    \begin{split}
        \mathbf{T}_{\text{sym}} =& - \frac{kE_i}{\rho_f} c_f\Im{\big[2\pmb{\alpha}_{vv}\mathbf{e}_z\big]}\times\mathbf{e}_z\sin^2(kz),\\
        \mathbf{T}_{\text{asym}} =& \frac{kE_i}{\rho_f}\Big(\frac{1}{\kappa_f}\Re{\big[\pmb{\alpha}_{vp}\big]}\times\mathbf{e}_z\Big) \sin(2kz),
    \end{split}
    \label{eq:torque_exp_PSW}
\end{align}
where subscript sym denotes the partial force and torque associated with direct-polarization coefficients $\alpha_{pp}$ and $\pmb{\alpha}_{vv}$ \cite{af_bruus7,shahrokh2021_WC_ARF}.
The terms with the subscript asym are induced by the Willis coupling coefficients $\pmb{\alpha}_{pv}$ and $\pmb{\alpha}_{vp}$, which obey $\pmb{\alpha}_{pv}=j\omega\rho_f\pmb{\alpha}_{vp}^T$ for reciprocal media.
These terms show the correlation between shape asymmetry and the two fields of acoustic radiation force and torque.
This is illustrated in Fig.~\ref{fig:schematic_sph_MA}(b), with objects exhibiting strong Willis coupling being trapped in a different location than their symmetric counterparts, cf. Fig.~\ref{fig:schematic_sph_MA}(a).
\textcolor{black}{Eqs.~\eqref{eq:force_exp_PSW} and  \eqref{eq:torque_exp_PSW} were derived using Gorkov's far-field approach, from the radiated momentum and the monopole-dipole approximation of the scattered pressure field \cite{shahrokh2021_WC_ARF}, and are applicable to objects with arbitrary shape in a plane standing wave.
Similar expressions for radiation force and torque for other types of incident waves, e.g. Bessel, Gaussian and Vortex beams, can be derived  from Eqs.~(17) and (24) in \cite{shahrokh2021_WC_ARF}.
The polarizability coefficients are calculated numerically using the novel approach in \cite{shahrokh2021_WC_ARF}, which is based on the Boundary Element Method and multipole translation and rotation theory.
Details of these numerical calculations and the verification study are provided in \cite{shahrokh2021_WC_ARF}.
Similar to the torque expression for spheroids, Eq.~(21) in \cite{leao2020_spheroid}, $T_{\text{sym}}$ in Eq.~\eqref{eq:torque_exp_PSW}, and Eq.~(I.9) in \cite{supp_notes}, show the spatial dependences of $\sin^2(kz)$ and $\sin(2\theta)$ that apply to objects of arbitrary shape in a plane standing wave.} 


An incident pressure field of 10\,mm wavelength in air is considered, with the speed of sound $c_f = 343.140$\,$\text{ms}^{-1}$ and mean density $\rho_f = 1.204$\,$\text{kgm}^{-3}$\textcolor{black}{, to investigate force reversal in airborne levitation within the long wavelength range $ka < 1$.
} 
\textcolor{black}{
The acoustic radiation force and torque values are normalized as force contrast factor $\mathbf{Q} =  \mathbf{F}/\big(E_i \pi a^2\big)$ and torque contrast factor $\mathbf{Z} = \mathbf{T}/\big(E_i \pi a^3\big)$.} 
The energy density of the incident wave is $E_i=p_i^2/(4\rho_f c_f^2)$.
\textcolor{black}{
Although the effects of viscous losses can be captured by the polarizability coefficients, they are neglected since the depth of the viscous boundary layer is at least three times smaller than the smallest $W$ in this study \textcolor{black}{(for validation see the Supplementary Notes \cite{supp_notes})}.
}

\mathchardef\mhyphen="2D
\begin{figure*}[ht!]
    \psfrag{(La)}[lB][lc][1.1]{(a)}
    \psfrag{aA1}[c][c][0.9]{$ka_{res}$}
    \psfrag{aA2}[c][c][0.9]{\textcolor{red}{$ka_{root}$}}
    \psfrag{ax}[c][c][1.1]{$ka$}
    \psfrag{ax1}[c][c][1.1]{$10^{\mhyphen1}$}
    \psfrag{ax2}[c][c][1.1]{$10^{0}$}
    \psfrag{ay}[B][l][1.1]{$\vert \overline{\alpha}\vert / \big(6\pi/c_f^2k^3\big)$}
    \psfrag{ay1}[Br][cr][1.1]{$10^{\mhyphen10}$}
    \psfrag{ay2}[r][r][1.1]{$10^{\mhyphen5}$}
    \psfrag{ay3}[r][r][1.1]{$10^{0}$}
    \psfrag{at}[l][l][0.8]{}
    \psfrag{al1}[l][l][0.8]{$\overline{\alpha}_{pp}$}
    \psfrag{al2}[l][l][0.8]{$\overline{\alpha}_{pv}^{z}$}
    \psfrag{al3}[l][l][0.8]{$\overline{\alpha}_{vp}^{z}$}
    \psfrag{al4}[l][l][0.8]{$\overline{\alpha}_{vv}^{xx}$}
    \psfrag{al5}[l][l][0.8]{$\overline{\alpha}_{vv}^{yy}$}
    \psfrag{al6}[l][l][0.8]{$\overline{\alpha}_{vv}^{zz}$}
    \psfrag{(Lb)}[lB][lc][1.1]{(b)}
    \psfrag{bx}[c][c][1.1]{$ka$}
    \psfrag{bx1}[c][c][1.1]{$10^{\mhyphen1}$}
    \psfrag{bx2}[c][c][1.1]{$10^{0}$}
    \psfrag{by}[B][c][1.1]{$\vert Q_x\vert, \vert Q_z\vert$}
    \psfrag{by1}[Br][cr][1.1]{$10^{\mhyphen3}$}
    \psfrag{by2}[r][r][1.1]{$10^{\mhyphen2}$}
    \psfrag{by3}[r][r][1.1]{$10^{\mhyphen1}$}
    \psfrag{by4}[c][c][1.1]{$10^{0}\quad$}
    \psfrag{by5}[c][c][1.1]{$10^{1}\quad$}
    \psfrag{bd1}[l][l][0.8]{numerical}
    \psfrag{bd2}[l][l][0.8]{sym$+$asym}
    \psfrag{bd3}[l][l][0.8]{sym}
    \psfrag{bd4}[l][l][0.8]{asym}
    \psfrag{ba1}[l][l][0.8]{$ka_{\text{(II)}}$=$0.157$}
    \psfrag{ba2}[l][l][0.9]{$ka_{\text{(I)}}$=$0.094$}
    \psfrag{ba3}[l][l][0.9]{$ka_{\text{(III)}}$=$0.176$}
    \psfrag{(Lc)}[lB][lc][1.1]{(c)}
    \psfrag{cx}[c][c][1.1]{$ka$}
    \psfrag{cx1}[c][c][1.1]{$10^{\mhyphen1}$}
    \psfrag{cx2}[c][c][1.1]{$10^{0}$}
    \psfrag{cy}[B][c][1.1]{$\vert Z_y\vert$}
    \psfrag{cy1}[Br][cr][1.1]{$10^{\mhyphen3}$}
    \psfrag{cy2}[r][r][1.1]{$10^{\mhyphen2}$}
    \psfrag{cy3}[r][r][1.1]{$10^{\mhyphen1}$}
    \psfrag{cy4}[c][c][1.1]{$10^{0}\quad$}
    \psfrag{cy5}[c][c][1.1]{$10^{1}\quad$}
    \psfrag{ca1}[r][r][0.9]{$\theta$=$\pi/4$}
    \psfrag{ca2}[l][l][0.9]{\textcolor{red}{$\hat{x}$}}
    \psfrag{ca3}[lc][lt][0.9]{$x$}
    \psfrag{ca4}[l][l][0.9]{\textcolor{red}{$\hat{z}$}}
    \psfrag{ca5}[lc][lt][0.9]{$z$}
    \psfrag{(Ld)}[lB][lc][1.1]{\textcolor{white}{(d)}}
    \psfrag{dx}[c][c][1.1]{$z/\lambda$}
    \psfrag{dx1}[c][c][1.1]{$0$}
    \psfrag{dx2}[c][c][1.1]{$0.25$}
    \psfrag{dx3}[c][c][1.1]{$0.5$}
    \psfrag{dy}[c][c][1.1]{Incidence angle $\theta$}
    \psfrag{dy1}[Br][cr][1.1]{$0$}
    \psfrag{dy2}[r][r][1.1]{$90$}
    \psfrag{dy3}[r][r][1.1]{$180$}
    \psfrag{dy4}[r][r][1.1]{$270$}
    \psfrag{dy5}[r][r][1.1]{$360$}
    \psfrag{dc1}[l][l][1.1]{$\mhyphen 0.05$}
    \psfrag{dc2}[l][l][1.1]{$0$}
    \psfrag{dc3}[l][l][1.1]{$0.05$}
    \psfrag{dt}[c][c][1.1]{Norm. force $Q_{\hat{z}}$, $ka_{\text{(II)}}$}
    \psfrag{(Le)}[lB][lc][1.1]{\textcolor{white}{(e)}}
    \psfrag{ex}[c][c][1.1]{$z/\lambda$}
    \psfrag{ex1}[c][c][1.1]{$0$}
    \psfrag{ex2}[c][c][1.1]{$0.25$}
    \psfrag{ex3}[c][c][1.1]{$0.5$}
    \psfrag{ey}[c][c][1.1]{}
    \psfrag{ey1}[Br][cr][1.1]{}
    \psfrag{ey2}[r][r][1.1]{}
    \psfrag{ey3}[r][r][1.1]{}
    \psfrag{ey4}[r][r][1.1]{}
    \psfrag{ey5}[r][r][1.1]{}
    \psfrag{ec1}[l][l][1.1]{$\mhyphen 0.3$}
    \psfrag{ec2}[l][l][1.1]{$\mhyphen 0.15$}
    \psfrag{ec3}[l][l][1.1]{$0$}
    \psfrag{ec4}[l][l][1.1]{$0.15$}
    \psfrag{ec5}[l][l][1.1]{$0.3$}
    \psfrag{et}[c][c][1.1]{Norm. torque $Z_{y}$, $ka_{\text{(II)}}$}
    \psfrag{(Lf)}[lB][lc][1.1]{(f)}
    \psfrag{fx}[c][c][1.1]{$z/\lambda$}
    \psfrag{fx1}[c][c][1.1]{$0$}
    \psfrag{fx2}[c][c][1.1]{$0.25$}
    \psfrag{fx3}[c][c][1.1]{$0.5$}
    \psfrag{fy}[c][c][1.1]{}
    \psfrag{fy1}[Br][cr][1.1]{}
    \psfrag{fy2}[r][r][1.1]{}
    \psfrag{fy3}[r][r][1.1]{}
    \psfrag{fy4}[r][r][1.1]{}
    \psfrag{fy5}[r][r][1.1]{}
    \psfrag{fd1}[l][l][0.9]{zero force}
    \psfrag{fd2}[l][l][0.9]{zero torque}
    \psfrag{fd3}[l][l][0.9]{stable trap}
    \psfrag{fd4}[l][l][0.9]{unstable trap}
    \psfrag{fd5}[l][l][0.9]{saddle point}
    \psfrag{PN}[c][c][1]{\textcolor{red}{P.N.}}
    \psfrag{MA}[l][l][1]{\textcolor{blue}{Meta-atom}}
    \psfrag{ES}[l][l][1]{\textcolor{red}{Equiv. Sph.}}
    \psfrag{EP}[l][l][1]{\textcolor{red}{Equiv. Prolate}}
    \psfrag{numerical}[c][c][0.9]{numerical}
    \psfrag{sym}[c][c][0.9]{sym}
    \psfrag{asym}[c][c][0.9]{asym}
    \includegraphics[width=0.9\textwidth]{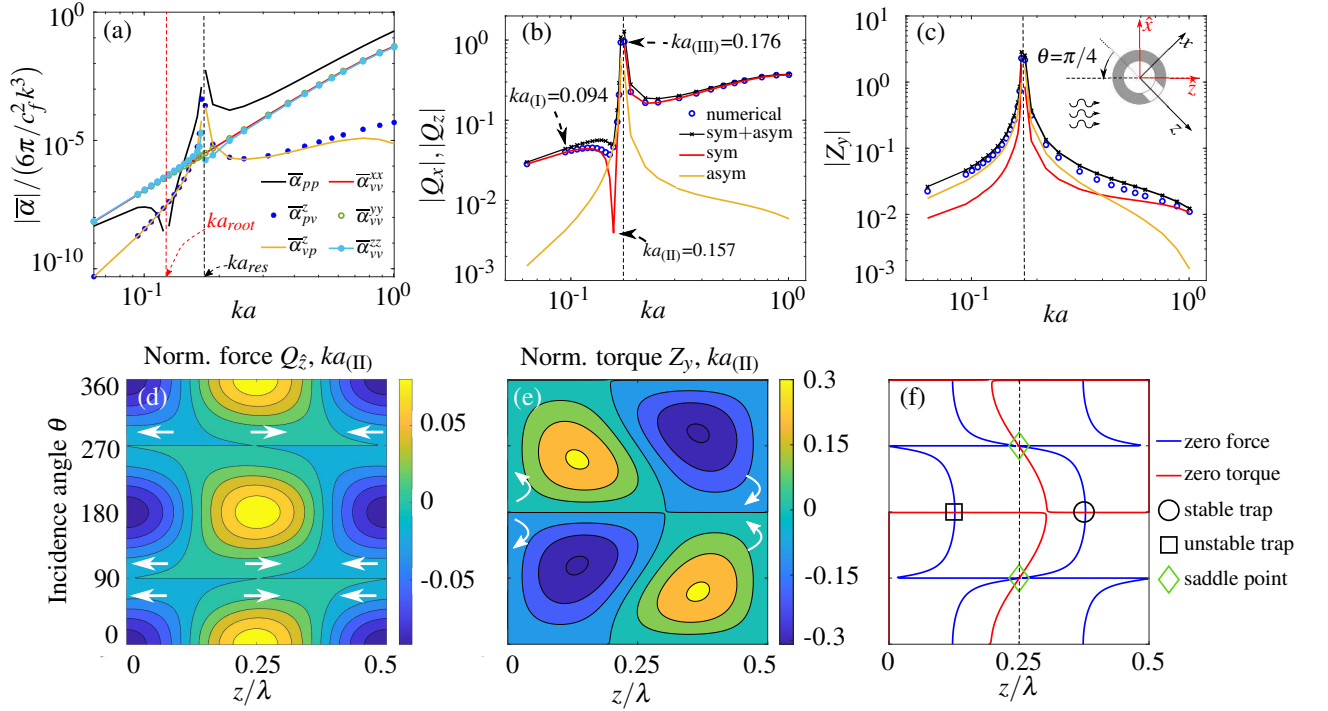}
    \caption{Panel (a) shows the absolute value of the non-zero and normalized polarizability coefficients \textcolor{black}{$\pmb{\alpha}$}, for the 3D meta-atom.
    Normalized radiation force and torque with respect to the size index $ka$ for the $\theta=-\pi/4$ incidence angle are shown in panels (b) and (c), respectively. 
    The magnitude of these quantities are denoted by $\vert\mathbf{Q}\vert$ and $\vert\mathbf{Z}\vert$.
    Panels (d) and (e) are the normalized force and torque, respectively, for the case of $ka_{\text{(II)}} = 0.157$ with respect to position \textcolor{black}{$z/\lambda$} and incidence angle \textcolor{black}{$\theta$.
    The arrows indicate the directions of the force and torque in different regions of panels (d) and (e).}
    Panel (f) shows the loci of zero force and zero torque and locations of saddle points, unstable and stable acoustic traps.
    \textcolor{black}{Numerical force and torque results are obtained using direct integration of stresses over a fictitious surface enclosing the object.}}
    \label{fig:Qxz_Zy_sph_1_pos}
\end{figure*}

The non-zero coefficients of the normalized polarizability tensor of the meta-atom are shown in Fig.~\ref{fig:Qxz_Zy_sph_1_pos}(a), with respect to the size index $ka$.
The results are scaled by $6\pi/(c_f^2k^3)$, which is the maximum permissible value of normalized Willis coupling coefficients \cite{Alu2018maxWC}. 
Real and imaginary parts of polarizability coefficients are shown in Fig.~(2) of the Supplementary Notes \cite{supp_notes}.
A resonant peak is found at $ka\approx 0.17$, and a trough, indicating a zero root, is observed for $\alpha_{pp}$ at $ka\approx 0.12$, implying no contribution from the incident pressure field to the monopole moment.
The resonant peak is only observed in the polarizability coefficients that are associated with the aperture facing the $z$-direction.
These two effects originate from the internal cavity, changing the scattering strength of the monopole mode.
\textcolor{black}{
The radiation force and torque obtained using Eqs.~\eqref{eq:force_exp_PSW} and \eqref{eq:torque_exp_PSW} are shown with respect $ka$ in Fig.~\ref{fig:Qxz_Zy_sph_1_pos}(b) and Fig.\ref{fig:Qxz_Zy_sph_1_pos}(c), respectively.
The meta-atom is oriented at a fixed angle of $-\pi/4$ with respect to the positive wave direction $\hat{z}$, as shown in the inset of Fig.\ref{fig:Qxz_Zy_sph_1_pos}(c), to ensure non-zero torque in the $y$-direction.
The force has two equal components in $x$- and $z$-directions.
Equations \eqref{eq:force_exp_PSW} and \eqref{eq:torque_exp_PSW} are verified by comparing with the force obtained by direct integration of radiation stresses over a fictitious surface of radius $4\lambda$ enclosing the meta-atom \cite{af_bruus7, FBW2015spheroid, shahrokh2021_WC_ARF}.} 
The contributions of the direct and Willis-coupling partial forces are plotted in Fig.~\ref{fig:Qxz_Zy_sph_1_pos}(b).
\textcolor{black}{
The local minimum in $Q_{\text{sym}}$ at $ka\approx 0.16$ emerges due to the root of $\alpha_{pp}$ at $ka_{\text{root}}\approx 0.12$,
and $\alpha_{vv}^{zz}$ being non-zero within this entire range of $ka$.} 
The radiation torque exhibits no local minimum since it is independent of the monopole scattering term $\alpha_{pp}$, as shown in Eq.~\eqref{eq:torque_exp_PSW}.
The orientation of the object with respect to the wave propagation direction changes the acoustic radiation force and radiation torque, as shown analytically in \textcolor{red}{\cite{supp_notes}}.

\textcolor{black}{Figures~\ref{fig:Qxz_Zy_sph_1_pos}(d) and \ref{fig:Qxz_Zy_sph_1_pos}(e) show the acoustic radiation force and radiation torque with respect to position $z$ and incidence angle $\theta$ for the specific case of $ka_{(\text{II})} \approx 0.16$, at which the radiation force is dominated by the Willis-coupling partial force.}  
The direction of the force and torque are indicated by the arrows in each region.
\textcolor{black}{In contrast to the case of symmetric objects, the radiation force is no longer zero at the pressure node $z/\lambda = 0.25$; implying a significant shift in the stable trapping location due to shape asymmetry.}
It is also observed that the force depends on angle $\theta$,
meaning that the stable acoustic trap occurs where both the force and torque are zero, corresponding to the intersection of the zero contours in Fig.~\ref{fig:Qxz_Zy_sph_1_pos}(d) and (e).
These zero force and torque contours are plotted in Fig.~\ref{fig:Qxz_Zy_sph_1_pos}(f), leading to four intersection points. 
\textcolor{black}{Only one of these points is stable (circle marker), since both the force and torque gradients are negative.}  
Analysis of the force and torque gradients reveals that the other three points are unstable.
\textcolor{black}{For the point marked by a square, any perturbation of position or orientation causes the particle to escape.}  
The points marked with diamonds are saddle points, where particular combinations of torque and force perturbation will lead to the particle escaping. 


\begin{figure*}
    \centering
    \psfrag{La}[lB][lc][1.1]{(a) $ka_{\text{(I)}}$}
    \psfrag{ax}[t][c][1.1]{}
    \psfrag{ax1}[c][c][1.1]{}
    \psfrag{ax2}[c][c][1.1]{}
    \psfrag{ax3}[c][c][1.1]{}
    \psfrag{ay}[c][c][1.1]{$Q_x$, $Q_z$}
    \psfrag{ay1}[Br][cr][1.1]{$-0.05$}
    \psfrag{ay2}[r][r][1.1]{$0$}
    \psfrag{ay3}[r][r][1.1]{$0.05$}
    \psfrag{Lb}[lB][lc][1.1]{(b) $ka_{\text{(II)}}$}
    \psfrag{bx}[t][c][1.1]{}
    \psfrag{bx1}[c][c][1.1]{}
    \psfrag{bx2}[c][c][1.1]{}
    \psfrag{bx3}[c][c][1.1]{}
    \psfrag{by}[b][c][1.1]{}
    \psfrag{by1}[Br][cr][1.1]{$-0.05$}
    \psfrag{by2}[r][r][1.1]{$0$}
    \psfrag{by3}[r][r][1.1]{$0.05$}
    \psfrag{Lc}[lB][lc][1.1]{(c) $ka_{\text{(III)}}$}
    \psfrag{cx}[t][c][1.1]{}
    \psfrag{cx1}[c][c][1.1]{}
    \psfrag{cx2}[c][c][1.1]{}
    \psfrag{cx3}[c][c][1.1]{}
    \psfrag{cy}[b][c][1.1]{}
    \psfrag{cy1}[r][r][1.1]{$-1$}
    \psfrag{cy2}[r][r][1.1]{$0$}
    \psfrag{cy3}[r][r][1.1]{$1$}
    \psfrag{Ld}[lB][lc][1.1]{(d) $ka_{\text{(I)}}$}
    \psfrag{dx}[c][c][1.1]{$z/\lambda$}
    \psfrag{dx1}[c][c][1.1]{$0$}
    \psfrag{dx2}[c][c][1.1]{$0.25$}
    \psfrag{dx3}[c][c][1.1]{$0.5$}
    \psfrag{dy}[c][c][1.1]{$Z_y$}
    \psfrag{dy1}[Br][cr][1.1]{$-0.05$}
    \psfrag{dy2}[r][r][1.1]{$0$}
    \psfrag{dy3}[r][r][1.1]{$0.05$}
    \psfrag{Le}[lB][lc][1.1]{(e) $ka_{\text{(II)}}$}
    \psfrag{ex}[c][c][1.1]{$z/\lambda$}
    \psfrag{ex1}[c][c][1.1]{$0$}
    \psfrag{ex2}[c][c][1.1]{$0.25$}
    \psfrag{ex3}[c][c][1.1]{$0.5$}
    \psfrag{ey}[b][c][1.1]{}
    \psfrag{ey1}[Br][cr][1.1]{$-0.3$}
    \psfrag{ey2}[r][r][1.1]{$0$}
    \psfrag{ey3}[r][r][1.1]{$0.3$}
    \psfrag{Lf}[lB][lc][1.1]{(f) $ka_{\text{(III)}}$}
    \psfrag{fx}[c][c][1.1]{$z/\lambda$}
    \psfrag{fx1}[c][c][1.1]{$0$}
    \psfrag{fx2}[c][c][1.1]{$0.25$}
    \psfrag{fx3}[c][c][1.1]{$0.5$}
    \psfrag{fy}[b][c][1.1]{}
    \psfrag{fy1}[Br][cr][1.1]{$-3$}
    \psfrag{fy2}[r][r][1.1]{$0$}
    \psfrag{fy3}[r][r][1.1]{$3$}
    \psfrag{PN}[c][c][0.9]{\textcolor{red}{P.N.}}
    \psfrag{MA}[l][l][0.9]{\textcolor{blue}{Meta-atom}}
    \psfrag{ES}[l][l][0.9]{\textcolor{red}{Equiv. Sph.}}
    \psfrag{EP}[l][l][0.9]{\textcolor{red}{Equiv. Prolate}}
    \psfrag{numerical}[c][c][0.9]{numerical}
    \psfrag{sym}[c][c][0.9]{sym}
    \psfrag{asym}[c][c][0.9]{asym}
    \includegraphics[width=\textwidth]{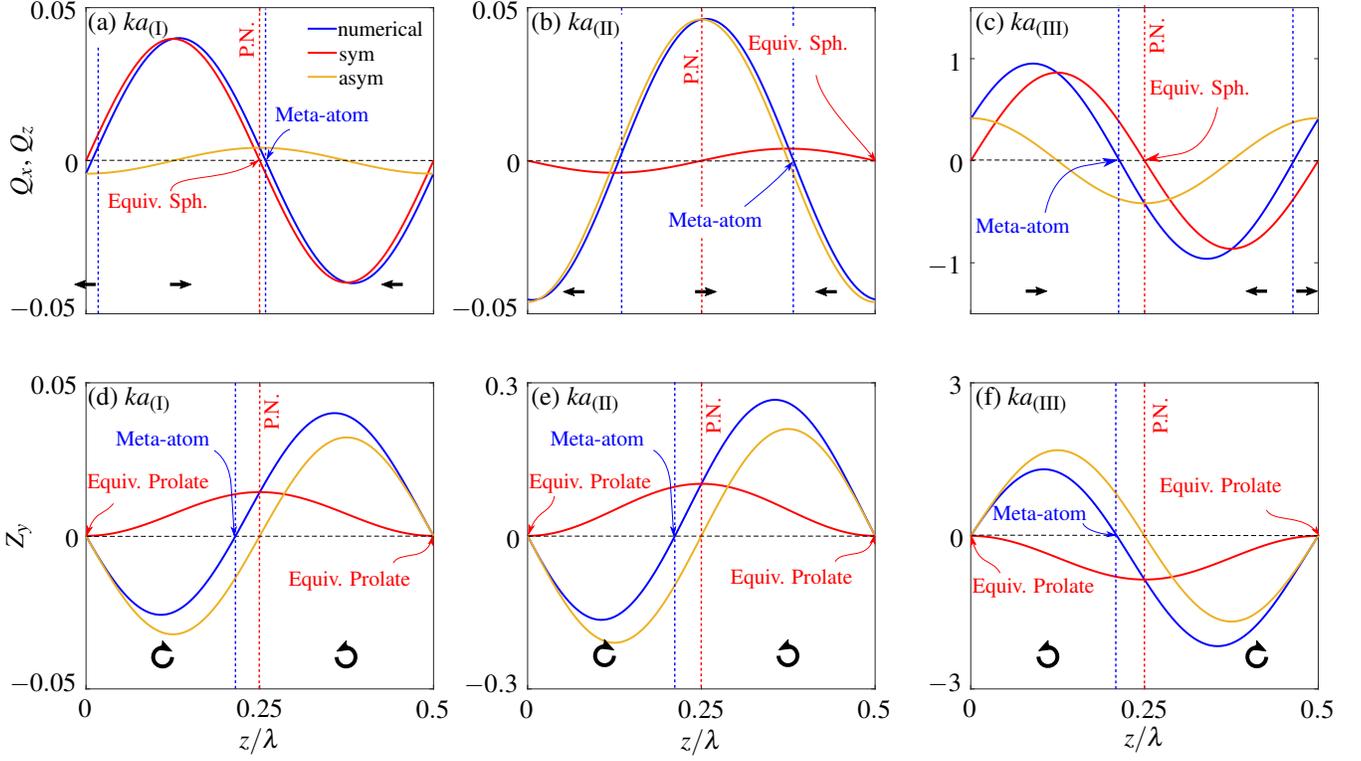}
    \caption{Variations of the normalized radiation force, panels (a)-(c), and radiation torque, panels (d)-(f), over a half wavelength span between two velocity nodes of a plane standing wave in the $z$ direction.
    The results are shown for \textcolor{black}{ $ka_{\text{(I)}} = 0.094$ (Rayleigh Regime), $ka_{\text{(II)}} = 0.157$ (minimum $F_{sym}$), and $ka_{\text{(III)}} = 0.176$ (resonance peak), and the orientation angle $\theta = -\pi/4$.
    \textcolor{black}{The equivalent spheroid is obtained from the polarizability tensor of the meta-atom by setting the Willis couplings to zero.
    For the equivalent sphere,  $\alpha^{zz}_{vv}$ is additionally set to $\alpha^{xx}_{vv}$.
    The trap locations of the equivalent sphere and spheroid are found from the analytical expressions in \cite{af_bruus7, Marston} and \cite{fan2008torque, leao2020_spheroid}, respectively.}} }
    \label{fig:QZ_reversal_size}
\end{figure*}

We now investigate the force and torque distributions at different positions \textcolor{black}{$z/\lambda$} within the standing wave, as shown in Fig.~\ref{fig:QZ_reversal_size}. 
\textcolor{black}{
A sphere and a prolate spheroid with similar polarizability response to the meta-atom are considered as reference cases, with their acoustic trap being at the pressure node \cite{af_bruus7, Marston, fan2008torque, leao2020_spheroid, shahrokh2021_WC_ARF}.
We quantify the role of Willis coupling as how far the trap location of the meta-atom shifts away from the pressure node.} 
\textcolor{black}{
Although equivalent objects are selected based on the polarizability response of the meta-atom, the prediction of their trap location is independent of their size as long as $ka\ll1$.} 
\textcolor{black}{Three} different frequency regimes \textcolor{black}{are considered} to show the drastic impact of \textcolor{black}{shape asymmetry} on the trapping location.
\textcolor{black}{Figures 3(a) and 3(d) show the results for $ka_{(\text{I})} =0.094$, corresponding to a low frequency limit where the Willis coupling is much smaller than the direct polarizability terms.}  
As expected, the Willis-coupling partial force (asym in Fig.~\ref{fig:QZ_reversal_size}(a)) is negligible compared to the direct polarization contribution, $F_{\text{sym}}$ (sym in Fig.~\ref{fig:QZ_reversal_size}(a)).
\textcolor{black}{Since $F_{\text{sym}}$ indicates the net force applied to the equivalent sphere, its trapping point is at the pressure node, where the force is zero with a negative gradient (Fig.~\ref{fig:QZ_reversal_size}(a)), matching the results of the classical theory for sound-hard spheres in a plane standing wave \cite{King_34,Gorkov_62,Yosioka_55,af_bruus7,Bruus_2012,shahrokh2020PRE,Sh2020_agglo}.} 
\textcolor{black}{However, the trap location of the meta-atom determined by the sum of the two partial forces, shown by the solid blue line, is slightly shifted from the pressure node, as a result of  $F_{\text{asym}}$ being dominant in that region.}  
\textcolor{black}{
In Fig.~\ref{fig:QZ_reversal_size}(d), we observe the Willis-coupling torque being dominant, resulting in an additional zero-torque location  at $z/\lambda \approx 0.21$, unlike the reference prolate spheroid that only experiences zero-torque at the velocity nodes $z/\lambda  = 0$ and $z/\lambda =0.5$ \cite{fan2008torque, leao2020_spheroid, shahrokh2021_WC_ARF}.} 

In Fig.~\ref{fig:QZ_reversal_size}(b) and \ref{fig:QZ_reversal_size}(e), we consider the frequency $ka_{(\text{II})}=0.157$, corresponding to a local minimum of the direct force term $F_\text{{sym}}$, so that the total radiation force is determined by the Willis-coupling contribution.
\textcolor{black}{This causes the stable trap to shift from the pressure node by $z/\lambda \approx 0.125$, to where  the symmetric contribution to the force is maximum - illustrating that asymmetric particles can show behavior directly contradicting the conventional Gorkov theory.} 
Although $F_\text{{sym}}$ is relatively small, we observe a direction reversal of the force vector in comparison to $ka_{(\text{I})}$.
Such force reversal is typically expected only for $ka>1$ \textcolor{black}{\cite{mitri2009Bessel, mitri2015ellCyl, zhang2018_reversals, zhang2021_phase_bessel}}, due to the resonant modes of the outer fluid domain.
However, our proposed meta-atom undergoes such force reversal due to the internal cavity acting as a Helmholtz resonator and the maximized Willis coupling associated with its single aperture.
The \textcolor{black}{net} torque is determined by the large contribution from the Willis-coupling partial torque, and the stable angular equilibrium occurs at a point further to the left of the pressure node than for $ka_{(\text{I})}$.

Close to the resonance peak, at \textcolor{black}{$ka_{(\text{III})}=0.176$} , the total radiation force and $F_\text{{sym}}$ shown in Fig.~\ref{fig:QZ_reversal_size}(c) exhibit the same direction as predicted by Gorkov theory.
\textcolor{black}{We also observe a reversal of the Willis-coupling partial force $F_\text{{asym}}$, resulting in a large negative shift of more than $10a$ from the pressure node.} 
The reversal of the torque shown in Fig.~\ref{fig:QZ_reversal_size}(f) indicates its tendency to orient the aperture in the opposite direction to that obtained for $ka_{(\text{I})}$ and $ka_{(\text{II})}$.
The stable angular equilibrium is still found at $z/\lambda \approx 0.21$, due to the direction reversal of both partial torques.



Our results provide physical insight into the relationship between shape asymmetry and force and torque reversal, which we show here are also possible for $ka<1$. 
Willis coupling in general describes structure-inherent asymmetry, which can be related to geometrical anomalies of a trapped object.
In addition to anomalous scattering, meta-atoms with strong Willis coupling were shown to experience different acoustic radiation forces and radiation torques, indicating the potential to control these nonlinear acoustic phenomena through Willis coupling.
Applications such as particle sorting, could be tuned to process objects with similar shape features by generating traps at locations other than pressure/velocity nodes.
\textcolor{black}{An example is to switch frequency during a sorting process \cite{liu2011_freq_switch}, between resonant and off-resonant states, which can be used as a propulsion technique to drive micro/nano-bots between target locations in an acoustofluidic system.} 
\textcolor{black}{Further, single objects can be re-oriented, not only in air but also in other fluids, by adjusting the wavelength according to its asymmetric feature size to maximise the Willis coupling.} 
As these effects already take place within the sub-wavelength regime, target-specific ultrasound manipulation techniques for the control of biological cell\textcolor{black}{s, rafts of particles and small tissue ensembles or mesoscopic asymmetrical structures such as insect appendages could be designed by mixing in specifically engineered meta-objects.}  
This control of field quantities using metamaterial concepts for shape-dependent separation of heterogeneous suspension mixtures or adapting self-assembly processes with tailored acoustic beams is likely to provide a fresh approach to the field of acoustophoresis.

This research was supported by the Australian Research Council through Discovery Projects DP200101708 and DP200100358.

\bibliography{ARC_project_source}

\providecommand{\noopsort}[1]{}\providecommand{\singleletter}[1]{#1}%
\begin{thebibliography}{60}%
\makeatletter
\providecommand \@ifxundefined [1]{%
 \@ifx{#1\undefined}
}%
\providecommand \@ifnum [1]{%
 \ifnum #1\expandafter \@firstoftwo
 \else \expandafter \@secondoftwo
 \fi
}%
\providecommand \@ifx [1]{%
 \ifx #1\expandafter \@firstoftwo
 \else \expandafter \@secondoftwo
 \fi
}%
\providecommand \natexlab [1]{#1}%
\providecommand \enquote  [1]{``#1''}%
\providecommand \bibnamefont  [1]{#1}%
\providecommand \bibfnamefont [1]{#1}%
\providecommand \citenamefont [1]{#1}%
\providecommand \href@noop [0]{\@secondoftwo}%
\providecommand \href [0]{\begingroup \@sanitize@url \@href}%
\providecommand \@href[1]{\@@startlink{#1}\@@href}%
\providecommand \@@href[1]{\endgroup#1\@@endlink}%
\providecommand \@sanitize@url [0]{\catcode `\\12\catcode `\$12\catcode
  `\&12\catcode `\#12\catcode `\^12\catcode `\_12\catcode `\%12\relax}%
\providecommand \@@startlink[1]{}%
\providecommand \@@endlink[0]{}%
\providecommand \url  [0]{\begingroup\@sanitize@url \@url }%
\providecommand \@url [1]{\endgroup\@href {#1}{\urlprefix }}%
\providecommand \urlprefix  [0]{URL }%
\providecommand \Eprint [0]{\href }%
\providecommand \doibase [0]{https://doi.org/}%
\providecommand \selectlanguage [0]{\@gobble}%
\providecommand \bibinfo  [0]{\@secondoftwo}%
\providecommand \bibfield  [0]{\@secondoftwo}%
\providecommand \translation [1]{[#1]}%
\providecommand \BibitemOpen [0]{}%
\providecommand \bibitemStop [0]{}%
\providecommand \bibitemNoStop [0]{.\EOS\space}%
\providecommand \EOS [0]{\spacefactor3000\relax}%
\providecommand \BibitemShut  [1]{\csname bibitem#1\endcsname}%
\let\auto@bib@innerbib\@empty
\bibitem [{\citenamefont {Bruus}(2011)}]{af_bruus1}%
  \BibitemOpen
  \bibfield  {author} {\bibinfo {author} {\bibfnamefont {H.}~\bibnamefont
  {Bruus}},\ }\href@noop {} {\bibfield  {journal} {\bibinfo  {journal} {Lab on
  a Chip}\ }\textbf {\bibinfo {volume} {11}},\ \bibinfo {pages} {3742}
  (\bibinfo {year} {2011})}\BibitemShut {NoStop}%
\bibitem [{\citenamefont {Bruus}(2012{\natexlab{a}})}]{af_bruus7}%
  \BibitemOpen
  \bibfield  {author} {\bibinfo {author} {\bibfnamefont {H.}~\bibnamefont
  {Bruus}},\ }\href@noop {} {\bibfield  {journal} {\bibinfo  {journal} {Lab on
  a Chip}\ }\textbf {\bibinfo {volume} {12}},\ \bibinfo {pages} {1014}
  (\bibinfo {year} {2012}{\natexlab{a}})}\BibitemShut {NoStop}%
\bibitem [{\citenamefont {Lenshof}\ \emph {et~al.}(2012)\citenamefont
  {Lenshof}, \citenamefont {Magnusson},\ and\ \citenamefont
  {Laurell}}]{af_Laurell}%
  \BibitemOpen
  \bibfield  {author} {\bibinfo {author} {\bibfnamefont {A.}~\bibnamefont
  {Lenshof}}, \bibinfo {author} {\bibfnamefont {C.}~\bibnamefont {Magnusson}},\
  and\ \bibinfo {author} {\bibfnamefont {T.}~\bibnamefont {Laurell}},\
  }\href@noop {} {\bibfield  {journal} {\bibinfo  {journal} {Lab on a Chip}\
  }\textbf {\bibinfo {volume} {12}},\ \bibinfo {pages} {1210} (\bibinfo {year}
  {2012})}\BibitemShut {NoStop}%
\bibitem [{\citenamefont {Bruus}(2012{\natexlab{b}})}]{af_bruus10}%
  \BibitemOpen
  \bibfield  {author} {\bibinfo {author} {\bibfnamefont {H.}~\bibnamefont
  {Bruus}},\ }\href@noop {} {\bibfield  {journal} {\bibinfo  {journal} {Lab on
  a Chip}\ }\textbf {\bibinfo {volume} {12}},\ \bibinfo {pages} {1578}
  (\bibinfo {year} {2012}{\natexlab{b}})}\BibitemShut {NoStop}%
\bibitem [{\citenamefont {Dual}\ \emph {et~al.}(2012)\citenamefont {Dual},
  \citenamefont {Hahn}, \citenamefont {Leibacher}, \citenamefont {M{\"o}ller},
  \citenamefont {Schwarz},\ and\ \citenamefont {Wang}}]{af_dual2012}%
  \BibitemOpen
  \bibfield  {author} {\bibinfo {author} {\bibfnamefont {J.}~\bibnamefont
  {Dual}}, \bibinfo {author} {\bibfnamefont {P.}~\bibnamefont {Hahn}}, \bibinfo
  {author} {\bibfnamefont {I.}~\bibnamefont {Leibacher}}, \bibinfo {author}
  {\bibfnamefont {D.}~\bibnamefont {M{\"o}ller}}, \bibinfo {author}
  {\bibfnamefont {T.}~\bibnamefont {Schwarz}},\ and\ \bibinfo {author}
  {\bibfnamefont {J.}~\bibnamefont {Wang}},\ }\href@noop {} {\bibfield
  {journal} {\bibinfo  {journal} {Lab on a Chip}\ }\textbf {\bibinfo {volume}
  {12}},\ \bibinfo {pages} {4010} (\bibinfo {year} {2012})}\BibitemShut
  {NoStop}%
\bibitem [{\citenamefont {Wiklund}\ \emph {et~al.}(2012)\citenamefont
  {Wiklund}, \citenamefont {Green},\ and\ \citenamefont {Ohlin}}]{af_wiklund}%
  \BibitemOpen
  \bibfield  {author} {\bibinfo {author} {\bibfnamefont {M.}~\bibnamefont
  {Wiklund}}, \bibinfo {author} {\bibfnamefont {R.}~\bibnamefont {Green}},\
  and\ \bibinfo {author} {\bibfnamefont {M.}~\bibnamefont {Ohlin}},\
  }\href@noop {} {\bibfield  {journal} {\bibinfo  {journal} {Lab on a Chip}\
  }\textbf {\bibinfo {volume} {12}},\ \bibinfo {pages} {2438} (\bibinfo {year}
  {2012})}\BibitemShut {NoStop}%
\bibitem [{\citenamefont {Hartono}\ \emph {et~al.}(2011)\citenamefont
  {Hartono}, \citenamefont {Liu}, \citenamefont {Tan}, \citenamefont {Then},
  \citenamefont {Yung},\ and\ \citenamefont {Lim}}]{Lim_2011}%
  \BibitemOpen
  \bibfield  {author} {\bibinfo {author} {\bibfnamefont {D.}~\bibnamefont
  {Hartono}}, \bibinfo {author} {\bibfnamefont {Y.}~\bibnamefont {Liu}},
  \bibinfo {author} {\bibfnamefont {P.~L.}\ \bibnamefont {Tan}}, \bibinfo
  {author} {\bibfnamefont {X.~Y.~S.}\ \bibnamefont {Then}}, \bibinfo {author}
  {\bibfnamefont {L.-Y.~L.}\ \bibnamefont {Yung}},\ and\ \bibinfo {author}
  {\bibfnamefont {K.-M.}\ \bibnamefont {Lim}},\ }\href@noop {} {\bibfield
  {journal} {\bibinfo  {journal} {Lab on a Chip}\ }\textbf {\bibinfo {volume}
  {11}},\ \bibinfo {pages} {4072} (\bibinfo {year} {2011})}\BibitemShut
  {NoStop}%
\bibitem [{\citenamefont {Mohapatra}\ \emph {et~al.}(2018)\citenamefont
  {Mohapatra}, \citenamefont {Sepehrirahnama},\ and\ \citenamefont
  {Lim}}]{Lim_2018}%
  \BibitemOpen
  \bibfield  {author} {\bibinfo {author} {\bibfnamefont {A.~R.}\ \bibnamefont
  {Mohapatra}}, \bibinfo {author} {\bibfnamefont {S.}~\bibnamefont
  {Sepehrirahnama}},\ and\ \bibinfo {author} {\bibfnamefont {K.-M.}\
  \bibnamefont {Lim}},\ }\href@noop {} {\bibfield  {journal} {\bibinfo
  {journal} {Physical Review E}\ }\textbf {\bibinfo {volume} {97}},\ \bibinfo
  {pages} {053105} (\bibinfo {year} {2018})}\BibitemShut {NoStop}%
\bibitem [{\citenamefont {Augustsson}\ \emph {et~al.}(2012)\citenamefont
  {Augustsson}, \citenamefont {Magnusson}, \citenamefont {Nordin},
  \citenamefont {Lilja},\ and\ \citenamefont {Laurell}}]{Laurell2012}%
  \BibitemOpen
  \bibfield  {author} {\bibinfo {author} {\bibfnamefont {P.}~\bibnamefont
  {Augustsson}}, \bibinfo {author} {\bibfnamefont {C.}~\bibnamefont
  {Magnusson}}, \bibinfo {author} {\bibfnamefont {M.}~\bibnamefont {Nordin}},
  \bibinfo {author} {\bibfnamefont {H.}~\bibnamefont {Lilja}},\ and\ \bibinfo
  {author} {\bibfnamefont {T.}~\bibnamefont {Laurell}},\ }\href@noop {}
  {\bibfield  {journal} {\bibinfo  {journal} {Analytical chemistry}\ }\textbf
  {\bibinfo {volume} {84}},\ \bibinfo {pages} {7954} (\bibinfo {year}
  {2012})}\BibitemShut {NoStop}%
\bibitem [{\citenamefont {Bernassau}\ \emph {et~al.}(2014)\citenamefont
  {Bernassau}, \citenamefont {Glynne-Jones}, \citenamefont {Gesellchen},
  \citenamefont {Riehle}, \citenamefont {Hill},\ and\ \citenamefont
  {Cumming}}]{Hill_2014}%
  \BibitemOpen
  \bibfield  {author} {\bibinfo {author} {\bibfnamefont {A.}~\bibnamefont
  {Bernassau}}, \bibinfo {author} {\bibfnamefont {P.}~\bibnamefont
  {Glynne-Jones}}, \bibinfo {author} {\bibfnamefont {F.}~\bibnamefont
  {Gesellchen}}, \bibinfo {author} {\bibfnamefont {M.}~\bibnamefont {Riehle}},
  \bibinfo {author} {\bibfnamefont {M.}~\bibnamefont {Hill}},\ and\ \bibinfo
  {author} {\bibfnamefont {D.}~\bibnamefont {Cumming}},\ }\href@noop {}
  {\bibfield  {journal} {\bibinfo  {journal} {Ultrasonics}\ }\textbf {\bibinfo
  {volume} {54}},\ \bibinfo {pages} {268} (\bibinfo {year} {2014})}\BibitemShut
  {NoStop}%
\bibitem [{\citenamefont {Antfolk}\ \emph {et~al.}(2015)\citenamefont
  {Antfolk}, \citenamefont {Magnusson}, \citenamefont {Augustsson},
  \citenamefont {Lilja},\ and\ \citenamefont {Laurell}}]{antfolk2015_CTC}%
  \BibitemOpen
  \bibfield  {author} {\bibinfo {author} {\bibfnamefont {M.}~\bibnamefont
  {Antfolk}}, \bibinfo {author} {\bibfnamefont {C.}~\bibnamefont {Magnusson}},
  \bibinfo {author} {\bibfnamefont {P.}~\bibnamefont {Augustsson}}, \bibinfo
  {author} {\bibfnamefont {H.}~\bibnamefont {Lilja}},\ and\ \bibinfo {author}
  {\bibfnamefont {T.}~\bibnamefont {Laurell}},\ }\href@noop {} {\bibfield
  {journal} {\bibinfo  {journal} {Analytical chemistry}\ }\textbf {\bibinfo
  {volume} {87}},\ \bibinfo {pages} {9322} (\bibinfo {year}
  {2015})}\BibitemShut {NoStop}%
\bibitem [{\citenamefont {Marzo}\ \emph {et~al.}(2015)\citenamefont {Marzo},
  \citenamefont {Seah}, \citenamefont {Drinkwater}, \citenamefont {Sahoo},
  \citenamefont {Long},\ and\ \citenamefont {Subramanian}}]{DW2015}%
  \BibitemOpen
  \bibfield  {author} {\bibinfo {author} {\bibfnamefont {A.}~\bibnamefont
  {Marzo}}, \bibinfo {author} {\bibfnamefont {S.~A.}\ \bibnamefont {Seah}},
  \bibinfo {author} {\bibfnamefont {B.~W.}\ \bibnamefont {Drinkwater}},
  \bibinfo {author} {\bibfnamefont {D.~R.}\ \bibnamefont {Sahoo}}, \bibinfo
  {author} {\bibfnamefont {B.}~\bibnamefont {Long}},\ and\ \bibinfo {author}
  {\bibfnamefont {S.}~\bibnamefont {Subramanian}},\ }\href@noop {} {\bibfield
  {journal} {\bibinfo  {journal} {Nature communications}\ }\textbf {\bibinfo
  {volume} {6}},\ \bibinfo {pages} {8661} (\bibinfo {year} {2015})}\BibitemShut
  {NoStop}%
\bibitem [{\citenamefont {Wijaya}\ \emph {et~al.}(2016)\citenamefont {Wijaya},
  \citenamefont {Mohapatra}, \citenamefont {Sepehrirahnama},\ and\
  \citenamefont {Lim}}]{Lim_2016}%
  \BibitemOpen
  \bibfield  {author} {\bibinfo {author} {\bibfnamefont {F.~B.}\ \bibnamefont
  {Wijaya}}, \bibinfo {author} {\bibfnamefont {A.~R.}\ \bibnamefont
  {Mohapatra}}, \bibinfo {author} {\bibfnamefont {S.}~\bibnamefont
  {Sepehrirahnama}},\ and\ \bibinfo {author} {\bibfnamefont {K.-M.}\
  \bibnamefont {Lim}},\ }\href@noop {} {\bibfield  {journal} {\bibinfo
  {journal} {Microfluidics and Nanofluidics}\ }\textbf {\bibinfo {volume}
  {20}},\ \bibinfo {pages} {69} (\bibinfo {year} {2016})}\BibitemShut {NoStop}%
\bibitem [{\citenamefont {Marzo}\ and\ \citenamefont
  {Drinkwater}(2019)}]{DW2019}%
  \BibitemOpen
  \bibfield  {author} {\bibinfo {author} {\bibfnamefont {A.}~\bibnamefont
  {Marzo}}\ and\ \bibinfo {author} {\bibfnamefont {B.~W.}\ \bibnamefont
  {Drinkwater}},\ }\href@noop {} {\bibfield  {journal} {\bibinfo  {journal}
  {PNAS}\ }\textbf {\bibinfo {volume} {116}},\ \bibinfo {pages} {84} (\bibinfo
  {year} {2019})}\BibitemShut {NoStop}%
\bibitem [{\citenamefont {Polychronopoulos}\ and\ \citenamefont
  {Memoli}(2020)}]{memoli2020AcLevMeta}%
  \BibitemOpen
  \bibfield  {author} {\bibinfo {author} {\bibfnamefont {S.}~\bibnamefont
  {Polychronopoulos}}\ and\ \bibinfo {author} {\bibfnamefont {G.}~\bibnamefont
  {Memoli}},\ }\href {https://doi.org/10.1038/s41598-020-60978-4} {\bibfield
  {journal} {\bibinfo  {journal} {Scientific reports}\ }\textbf {\bibinfo
  {volume} {10}},\ \bibinfo {pages} {1} (\bibinfo {year} {2020})}\BibitemShut
  {NoStop}%
\bibitem [{\citenamefont {King}(1934)}]{King_34}%
  \BibitemOpen
  \bibfield  {author} {\bibinfo {author} {\bibfnamefont {L.~V.}\ \bibnamefont
  {King}},\ }\href@noop {} {\bibfield  {journal} {\bibinfo  {journal}
  {Proceedings of the Royal Society of London. Series A-Mathematical and
  Physical Sciences}\ }\textbf {\bibinfo {volume} {147}},\ \bibinfo {pages}
  {212} (\bibinfo {year} {1934})}\BibitemShut {NoStop}%
\bibitem [{\citenamefont {Yosioka}\ and\ \citenamefont
  {Kawasima}(1955)}]{Yosioka_55}%
  \BibitemOpen
  \bibfield  {author} {\bibinfo {author} {\bibfnamefont {K.}~\bibnamefont
  {Yosioka}}\ and\ \bibinfo {author} {\bibfnamefont {Y.}~\bibnamefont
  {Kawasima}},\ }\href@noop {} {\bibfield  {journal} {\bibinfo  {journal} {Acta
  Acustica united with Acustica}\ }\textbf {\bibinfo {volume} {5}},\ \bibinfo
  {pages} {167} (\bibinfo {year} {1955})}\BibitemShut {NoStop}%
\bibitem [{\citenamefont {Gorkov}(1962)}]{Gorkov_62}%
  \BibitemOpen
  \bibfield  {author} {\bibinfo {author} {\bibfnamefont {L.~P.}\ \bibnamefont
  {Gorkov}},\ }\href@noop {} {\bibfield  {journal} {\bibinfo  {journal} {Soviet
  Physics - Doklady}\ }\textbf {\bibinfo {volume} {6}},\ \bibinfo {pages} {773}
  (\bibinfo {year} {1962})}\BibitemShut {NoStop}%
\bibitem [{\citenamefont {Doinikov}(1994{\natexlab{a}})}]{Doinikov1994_JFM}%
  \BibitemOpen
  \bibfield  {author} {\bibinfo {author} {\bibfnamefont {A.~A.}\ \bibnamefont
  {Doinikov}},\ }\href@noop {} {\bibfield  {journal} {\bibinfo  {journal}
  {Journal of Fluid Mechanics}\ }\textbf {\bibinfo {volume} {267}},\ \bibinfo
  {pages} {1} (\bibinfo {year} {1994}{\natexlab{a}})}\BibitemShut {NoStop}%
\bibitem [{\citenamefont {Settnes}\ and\ \citenamefont
  {Bruus}(2012)}]{Bruus_2012}%
  \BibitemOpen
  \bibfield  {author} {\bibinfo {author} {\bibfnamefont {M.}~\bibnamefont
  {Settnes}}\ and\ \bibinfo {author} {\bibfnamefont {H.}~\bibnamefont
  {Bruus}},\ }\href@noop {} {\bibfield  {journal} {\bibinfo  {journal}
  {Physical Review E}\ }\textbf {\bibinfo {volume} {85}},\ \bibinfo {pages}
  {016327} (\bibinfo {year} {2012})}\BibitemShut {NoStop}%
\bibitem [{\citenamefont {Silva}\ and\ \citenamefont
  {Bruus}(2014)}]{silva_Bruus}%
  \BibitemOpen
  \bibfield  {author} {\bibinfo {author} {\bibfnamefont {G.~T.}\ \bibnamefont
  {Silva}}\ and\ \bibinfo {author} {\bibfnamefont {H.}~\bibnamefont {Bruus}},\
  }\href@noop {} {\bibfield  {journal} {\bibinfo  {journal} {Physical Review
  E}\ }\textbf {\bibinfo {volume} {90}},\ \bibinfo {pages} {063007} (\bibinfo
  {year} {2014})}\BibitemShut {NoStop}%
\bibitem [{\citenamefont {Marston}(2006)}]{marston2006axial}%
  \BibitemOpen
  \bibfield  {author} {\bibinfo {author} {\bibfnamefont {P.~L.}\ \bibnamefont
  {Marston}},\ }\href@noop {} {\bibfield  {journal} {\bibinfo  {journal} {The
  Journal of the Acoustical Society of America}\ }\textbf {\bibinfo {volume}
  {120}},\ \bibinfo {pages} {3518} (\bibinfo {year} {2006})}\BibitemShut
  {NoStop}%
\bibitem [{\citenamefont {Mitri}(2015{\natexlab{a}})}]{mitri2015}%
  \BibitemOpen
  \bibfield  {author} {\bibinfo {author} {\bibfnamefont {F.}~\bibnamefont
  {Mitri}},\ }\href@noop {} {\bibfield  {journal} {\bibinfo  {journal} {Wave
  Motion}\ }\textbf {\bibinfo {volume} {57}},\ \bibinfo {pages} {231} (\bibinfo
  {year} {2015}{\natexlab{a}})}\BibitemShut {NoStop}%
\bibitem [{\citenamefont {Garcia-Sabat{\'e}}\ \emph {et~al.}(2014)\citenamefont
  {Garcia-Sabat{\'e}}, \citenamefont {Castro}, \citenamefont {Hoyos},\ and\
  \citenamefont {Gonz{\'a}lez-Cinca}}]{garcia2014experimental}%
  \BibitemOpen
  \bibfield  {author} {\bibinfo {author} {\bibfnamefont {A.}~\bibnamefont
  {Garcia-Sabat{\'e}}}, \bibinfo {author} {\bibfnamefont {A.}~\bibnamefont
  {Castro}}, \bibinfo {author} {\bibfnamefont {M.}~\bibnamefont {Hoyos}},\ and\
  \bibinfo {author} {\bibfnamefont {R.}~\bibnamefont {Gonz{\'a}lez-Cinca}},\
  }\href@noop {} {\bibfield  {journal} {\bibinfo  {journal} {The Journal of the
  Acoustical Society of America}\ }\textbf {\bibinfo {volume} {135}},\ \bibinfo
  {pages} {1056} (\bibinfo {year} {2014})}\BibitemShut {NoStop}%
\bibitem [{\citenamefont {Doinikov}(1994{\natexlab{b}})}]{Doinikov1994_Proc}%
  \BibitemOpen
  \bibfield  {author} {\bibinfo {author} {\bibfnamefont {A.~A.}\ \bibnamefont
  {Doinikov}},\ }\href@noop {} {\bibfield  {journal} {\bibinfo  {journal}
  {Proceedings of the Royal Society of London. Series A: Mathematical and
  Physical Sciences}\ }\textbf {\bibinfo {volume} {447}},\ \bibinfo {pages}
  {447} (\bibinfo {year} {1994}{\natexlab{b}})}\BibitemShut {NoStop}%
\bibitem [{\citenamefont {Foresti}\ \emph {et~al.}(2012)\citenamefont
  {Foresti}, \citenamefont {Nabavi},\ and\ \citenamefont
  {Poulikakos}}]{foresti2012ellSpheroid}%
  \BibitemOpen
  \bibfield  {author} {\bibinfo {author} {\bibfnamefont {D.}~\bibnamefont
  {Foresti}}, \bibinfo {author} {\bibfnamefont {M.}~\bibnamefont {Nabavi}},\
  and\ \bibinfo {author} {\bibfnamefont {D.}~\bibnamefont {Poulikakos}},\
  }\href@noop {} {\bibfield  {journal} {\bibinfo  {journal} {Journal of Fluid
  Mechanics}\ }\textbf {\bibinfo {volume} {709}},\ \bibinfo {pages} {581}
  (\bibinfo {year} {2012})}\BibitemShut {NoStop}%
\bibitem [{\citenamefont {Wijaya}\ and\ \citenamefont
  {Lim}(2015)}]{FBW2015spheroid}%
  \BibitemOpen
  \bibfield  {author} {\bibinfo {author} {\bibfnamefont {F.~B.}\ \bibnamefont
  {Wijaya}}\ and\ \bibinfo {author} {\bibfnamefont {K.-M.}\ \bibnamefont
  {Lim}},\ }\href@noop {} {\bibfield  {journal} {\bibinfo  {journal} {Acta
  Acustica united with Acustica}\ }\textbf {\bibinfo {volume} {101}},\ \bibinfo
  {pages} {531} (\bibinfo {year} {2015})}\BibitemShut {NoStop}%
\bibitem [{\citenamefont {Mitri}(2015{\natexlab{b}})}]{mitri2015ellCyl}%
  \BibitemOpen
  \bibfield  {author} {\bibinfo {author} {\bibfnamefont {F.}~\bibnamefont
  {Mitri}},\ }\href@noop {} {\bibfield  {journal} {\bibinfo  {journal} {Journal
  of Applied Physics}\ }\textbf {\bibinfo {volume} {118}},\ \bibinfo {pages}
  {214903} (\bibinfo {year} {2015}{\natexlab{b}})}\BibitemShut {NoStop}%
\bibitem [{\citenamefont {Wei}\ \emph {et~al.}(2004)\citenamefont {Wei},
  \citenamefont {Thiessen},\ and\ \citenamefont {Marston}}]{wei2004cyl}%
  \BibitemOpen
  \bibfield  {author} {\bibinfo {author} {\bibfnamefont {W.}~\bibnamefont
  {Wei}}, \bibinfo {author} {\bibfnamefont {D.~B.}\ \bibnamefont {Thiessen}},\
  and\ \bibinfo {author} {\bibfnamefont {P.~L.}\ \bibnamefont {Marston}},\
  }\href@noop {} {\bibfield  {journal} {\bibinfo  {journal} {The Journal of the
  Acoustical Society of America}\ }\textbf {\bibinfo {volume} {116}},\ \bibinfo
  {pages} {201} (\bibinfo {year} {2004})}\BibitemShut {NoStop}%
\bibitem [{\citenamefont {Xie}\ and\ \citenamefont
  {Wei}(2004)}]{xie2004ARFdisc}%
  \BibitemOpen
  \bibfield  {author} {\bibinfo {author} {\bibfnamefont {W.}~\bibnamefont
  {Xie}}\ and\ \bibinfo {author} {\bibfnamefont {B.}~\bibnamefont {Wei}},\
  }\href@noop {} {\bibfield  {journal} {\bibinfo  {journal} {Physical Review
  E}\ }\textbf {\bibinfo {volume} {70}},\ \bibinfo {pages} {046611} (\bibinfo
  {year} {2004})}\BibitemShut {NoStop}%
\bibitem [{\citenamefont {Garbin}\ \emph {et~al.}(2015)\citenamefont {Garbin},
  \citenamefont {Leibacher}, \citenamefont {Hahn}, \citenamefont {Le~Ferrand},
  \citenamefont {Studart},\ and\ \citenamefont {Dual}}]{garbin2015ARFdisk}%
  \BibitemOpen
  \bibfield  {author} {\bibinfo {author} {\bibfnamefont {A.}~\bibnamefont
  {Garbin}}, \bibinfo {author} {\bibfnamefont {I.}~\bibnamefont {Leibacher}},
  \bibinfo {author} {\bibfnamefont {P.}~\bibnamefont {Hahn}}, \bibinfo {author}
  {\bibfnamefont {H.}~\bibnamefont {Le~Ferrand}}, \bibinfo {author}
  {\bibfnamefont {A.}~\bibnamefont {Studart}},\ and\ \bibinfo {author}
  {\bibfnamefont {J.}~\bibnamefont {Dual}},\ }\href@noop {} {\bibfield
  {journal} {\bibinfo  {journal} {The Journal of the Acoustical Society of
  America}\ }\textbf {\bibinfo {volume} {138}},\ \bibinfo {pages} {2759}
  (\bibinfo {year} {2015})}\BibitemShut {NoStop}%
\bibitem [{\citenamefont {Sepehrirahnama}\ and\ \citenamefont
  {Lim}(2020{\natexlab{a}})}]{shahrokh2020PRE}%
  \BibitemOpen
  \bibfield  {author} {\bibinfo {author} {\bibfnamefont {S.}~\bibnamefont
  {Sepehrirahnama}}\ and\ \bibinfo {author} {\bibfnamefont {K.-M.}\
  \bibnamefont {Lim}},\ }\href {https://doi.org/10.1103/PhysRevE.102.043307}
  {\bibfield  {journal} {\bibinfo  {journal} {Physical Review E}\ }\textbf
  {\bibinfo {volume} {102}},\ \bibinfo {pages} {043307} (\bibinfo {year}
  {2020}{\natexlab{a}})}\BibitemShut {NoStop}%
\bibitem [{\citenamefont {Quan}\ \emph {et~al.}(2018)\citenamefont {Quan},
  \citenamefont {Ra’di}, \citenamefont {Sounas},\ and\ \citenamefont
  {Al{\`u}}}]{Alu2018maxWC}%
  \BibitemOpen
  \bibfield  {author} {\bibinfo {author} {\bibfnamefont {L.}~\bibnamefont
  {Quan}}, \bibinfo {author} {\bibfnamefont {Y.}~\bibnamefont {Ra’di}},
  \bibinfo {author} {\bibfnamefont {D.~L.}\ \bibnamefont {Sounas}},\ and\
  \bibinfo {author} {\bibfnamefont {A.}~\bibnamefont {Al{\`u}}},\ }\href@noop
  {} {\bibfield  {journal} {\bibinfo  {journal} {Physical Review Letters}\
  }\textbf {\bibinfo {volume} {120}},\ \bibinfo {pages} {254301} (\bibinfo
  {year} {2018})}\BibitemShut {NoStop}%
\bibitem [{\citenamefont {Jordaan}\ \emph {et~al.}(2018)\citenamefont
  {Jordaan}, \citenamefont {Punzet}, \citenamefont {Melnikov}, \citenamefont
  {Sanches}, \citenamefont {Oberst}, \citenamefont {Marburg},\ and\
  \citenamefont {Powell}}]{jordaan2018}%
  \BibitemOpen
  \bibfield  {author} {\bibinfo {author} {\bibfnamefont {J.}~\bibnamefont
  {Jordaan}}, \bibinfo {author} {\bibfnamefont {S.}~\bibnamefont {Punzet}},
  \bibinfo {author} {\bibfnamefont {A.}~\bibnamefont {Melnikov}}, \bibinfo
  {author} {\bibfnamefont {A.}~\bibnamefont {Sanches}}, \bibinfo {author}
  {\bibfnamefont {S.}~\bibnamefont {Oberst}}, \bibinfo {author} {\bibfnamefont
  {S.}~\bibnamefont {Marburg}},\ and\ \bibinfo {author} {\bibfnamefont {D.~A.}\
  \bibnamefont {Powell}},\ }\href@noop {} {\bibfield  {journal} {\bibinfo
  {journal} {Applied Physics Letters}\ }\textbf {\bibinfo {volume} {113}},\
  \bibinfo {pages} {224102} (\bibinfo {year} {2018})}\BibitemShut {NoStop}%
\bibitem [{\citenamefont {Melnikov}\ \emph {et~al.}(2019)\citenamefont
  {Melnikov}, \citenamefont {Chiang}, \citenamefont {Quan}, \citenamefont
  {Oberst}, \citenamefont {Al{\`u}}, \citenamefont {Marburg},\ and\
  \citenamefont {Powell}}]{anton2019}%
  \BibitemOpen
  \bibfield  {author} {\bibinfo {author} {\bibfnamefont {A.}~\bibnamefont
  {Melnikov}}, \bibinfo {author} {\bibfnamefont {Y.~K.}\ \bibnamefont
  {Chiang}}, \bibinfo {author} {\bibfnamefont {L.}~\bibnamefont {Quan}},
  \bibinfo {author} {\bibfnamefont {S.}~\bibnamefont {Oberst}}, \bibinfo
  {author} {\bibfnamefont {A.}~\bibnamefont {Al{\`u}}}, \bibinfo {author}
  {\bibfnamefont {S.}~\bibnamefont {Marburg}},\ and\ \bibinfo {author}
  {\bibfnamefont {D.}~\bibnamefont {Powell}},\ }\href@noop {} {\bibfield
  {journal} {\bibinfo  {journal} {Nature communications}\ }\textbf {\bibinfo
  {volume} {10}},\ \bibinfo {pages} {1} (\bibinfo {year} {2019})}\BibitemShut
  {NoStop}%
\bibitem [{\citenamefont {Chiang}\ \emph {et~al.}(2020)\citenamefont {Chiang},
  \citenamefont {Oberst}, \citenamefont {Melnikov}, \citenamefont {Quan},
  \citenamefont {Marburg}, \citenamefont {Al{\`u}},\ and\ \citenamefont
  {Powell}}]{YK2020arraye}%
  \BibitemOpen
  \bibfield  {author} {\bibinfo {author} {\bibfnamefont {Y.~K.}\ \bibnamefont
  {Chiang}}, \bibinfo {author} {\bibfnamefont {S.}~\bibnamefont {Oberst}},
  \bibinfo {author} {\bibfnamefont {A.}~\bibnamefont {Melnikov}}, \bibinfo
  {author} {\bibfnamefont {L.}~\bibnamefont {Quan}}, \bibinfo {author}
  {\bibfnamefont {S.}~\bibnamefont {Marburg}}, \bibinfo {author} {\bibfnamefont
  {A.}~\bibnamefont {Al{\`u}}},\ and\ \bibinfo {author} {\bibfnamefont {D.~A.}\
  \bibnamefont {Powell}},\ }\href@noop {} {\bibfield  {journal} {\bibinfo
  {journal} {Physical Review Applied}\ }\textbf {\bibinfo {volume} {13}},\
  \bibinfo {pages} {064067} (\bibinfo {year} {2020})}\BibitemShut {NoStop}%
\bibitem [{\citenamefont {Ni}\ \emph {et~al.}(2019)\citenamefont {Ni},
  \citenamefont {Fang}, \citenamefont {Hou}, \citenamefont {Li},\ and\
  \citenamefont {Assouar}}]{ni2019metagratings}%
  \BibitemOpen
  \bibfield  {author} {\bibinfo {author} {\bibfnamefont {H.}~\bibnamefont
  {Ni}}, \bibinfo {author} {\bibfnamefont {X.}~\bibnamefont {Fang}}, \bibinfo
  {author} {\bibfnamefont {Z.}~\bibnamefont {Hou}}, \bibinfo {author}
  {\bibfnamefont {Y.}~\bibnamefont {Li}},\ and\ \bibinfo {author}
  {\bibfnamefont {B.}~\bibnamefont {Assouar}},\ }\href@noop {} {\bibfield
  {journal} {\bibinfo  {journal} {Physical Review B}\ }\textbf {\bibinfo
  {volume} {100}},\ \bibinfo {pages} {104104} (\bibinfo {year}
  {2019})}\BibitemShut {NoStop}%
\bibitem [{\citenamefont {Sieck}\ \emph {et~al.}(2017)\citenamefont {Sieck},
  \citenamefont {Al{\`u}},\ and\ \citenamefont {Haberman}}]{Alu2017WCorigin}%
  \BibitemOpen
  \bibfield  {author} {\bibinfo {author} {\bibfnamefont {C.~F.}\ \bibnamefont
  {Sieck}}, \bibinfo {author} {\bibfnamefont {A.}~\bibnamefont {Al{\`u}}},\
  and\ \bibinfo {author} {\bibfnamefont {M.~R.}\ \bibnamefont {Haberman}},\
  }\href@noop {} {\bibfield  {journal} {\bibinfo  {journal} {Physical Review
  B}\ }\textbf {\bibinfo {volume} {96}},\ \bibinfo {pages} {104303} (\bibinfo
  {year} {2017})}\BibitemShut {NoStop}%
\bibitem [{\citenamefont {Su}\ and\ \citenamefont {Norris}(2018)}]{norris2018}%
  \BibitemOpen
  \bibfield  {author} {\bibinfo {author} {\bibfnamefont {X.}~\bibnamefont
  {Su}}\ and\ \bibinfo {author} {\bibfnamefont {A.~N.}\ \bibnamefont
  {Norris}},\ }\href@noop {} {\bibfield  {journal} {\bibinfo  {journal}
  {Physical Review B}\ }\textbf {\bibinfo {volume} {98}},\ \bibinfo {pages}
  {174305} (\bibinfo {year} {2018})}\BibitemShut {NoStop}%
\bibitem [{\citenamefont {Melnikov}\ \emph {et~al.}(2020)\citenamefont
  {Melnikov}, \citenamefont {Maeder}, \citenamefont {Friedrich}, \citenamefont
  {Pozhanka}, \citenamefont {Wollmann}, \citenamefont {Scheffler},
  \citenamefont {Oberst}, \citenamefont {Powell},\ and\ \citenamefont
  {Marburg}}]{anton2020}%
  \BibitemOpen
  \bibfield  {author} {\bibinfo {author} {\bibfnamefont {A.}~\bibnamefont
  {Melnikov}}, \bibinfo {author} {\bibfnamefont {M.}~\bibnamefont {Maeder}},
  \bibinfo {author} {\bibfnamefont {N.}~\bibnamefont {Friedrich}}, \bibinfo
  {author} {\bibfnamefont {Y.}~\bibnamefont {Pozhanka}}, \bibinfo {author}
  {\bibfnamefont {A.}~\bibnamefont {Wollmann}}, \bibinfo {author}
  {\bibfnamefont {M.}~\bibnamefont {Scheffler}}, \bibinfo {author}
  {\bibfnamefont {S.}~\bibnamefont {Oberst}}, \bibinfo {author} {\bibfnamefont
  {D.}~\bibnamefont {Powell}},\ and\ \bibinfo {author} {\bibfnamefont
  {S.}~\bibnamefont {Marburg}},\ }\href
  {https://doi.org/https://doi.org/10.1121/10.0000857} {\bibfield  {journal}
  {\bibinfo  {journal} {The Journal of the Acoustical Society of America}\
  }\textbf {\bibinfo {volume} {147}},\ \bibinfo {pages} {1491} (\bibinfo {year}
  {2020})}\BibitemShut {NoStop}%
\bibitem [{\citenamefont {Jiang}\ \emph {et~al.}(2017)\citenamefont {Jiang},
  \citenamefont {Li},\ and\ \citenamefont {Zhang}}]{zhang2017_tv}%
  \BibitemOpen
  \bibfield  {author} {\bibinfo {author} {\bibfnamefont {X.}~\bibnamefont
  {Jiang}}, \bibinfo {author} {\bibfnamefont {Y.}~\bibnamefont {Li}},\ and\
  \bibinfo {author} {\bibfnamefont {L.}~\bibnamefont {Zhang}},\ }\href
  {https://doi.org/10.1121/1.4979682} {\bibfield  {journal} {\bibinfo
  {journal} {The Journal of the Acoustical Society of America}\ }\textbf
  {\bibinfo {volume} {141}},\ \bibinfo {pages} {EL363} (\bibinfo {year}
  {2017})}\BibitemShut {NoStop}%
\bibitem [{sup(2022)}]{supp_notes}%
  \BibitemOpen
  \bibfield  {title} {\bibinfo {title} {See supplemental material at [url will
  be inserted by publisher] for analytical derivation of the force and torque
  in terms of orientation angle, details of numerical modelling of
  polarizability coefficients, and verification study of the effects of viscous
  losses.},\ }\href@noop {} {\  (\bibinfo {year} {2022})}\BibitemShut {NoStop}%
\bibitem [{\citenamefont {Leao-Neto}\ \emph {et~al.}(2021)\citenamefont
  {Leao-Neto}, \citenamefont {Hoyos}, \citenamefont {Aider},\ and\
  \citenamefont {Silva}}]{leao2021_cyl_spheroid}%
  \BibitemOpen
  \bibfield  {author} {\bibinfo {author} {\bibfnamefont {J.~P.}\ \bibnamefont
  {Leao-Neto}}, \bibinfo {author} {\bibfnamefont {M.}~\bibnamefont {Hoyos}},
  \bibinfo {author} {\bibfnamefont {J.-L.}\ \bibnamefont {Aider}},\ and\
  \bibinfo {author} {\bibfnamefont {G.~T.}\ \bibnamefont {Silva}},\ }\href@noop
  {} {\bibfield  {journal} {\bibinfo  {journal} {The Journal of the Acoustical
  Society of America}\ }\textbf {\bibinfo {volume} {149}},\ \bibinfo {pages}
  {285} (\bibinfo {year} {2021})}\BibitemShut {NoStop}%
\bibitem [{\citenamefont {Le{\~a}o-Neto}\ \emph {et~al.}(2020)\citenamefont
  {Le{\~a}o-Neto}, \citenamefont {Lopes},\ and\ \citenamefont
  {Silva}}]{leao2020_spheroid}%
  \BibitemOpen
  \bibfield  {author} {\bibinfo {author} {\bibfnamefont {J.~P.}\ \bibnamefont
  {Le{\~a}o-Neto}}, \bibinfo {author} {\bibfnamefont {J.~H.}\ \bibnamefont
  {Lopes}},\ and\ \bibinfo {author} {\bibfnamefont {G.~T.}\ \bibnamefont
  {Silva}},\ }\href {https://doi.org/10.1121/10.0001016} {\bibfield  {journal}
  {\bibinfo  {journal} {The Journal of the Acoustical Society of America}\
  }\textbf {\bibinfo {volume} {147}},\ \bibinfo {pages} {2177} (\bibinfo {year}
  {2020})}\BibitemShut {NoStop}%
\bibitem [{\citenamefont {Marston}\ and\ \citenamefont
  {Zhang}(2017)}]{marston2017_phase_shift}%
  \BibitemOpen
  \bibfield  {author} {\bibinfo {author} {\bibfnamefont {P.~L.}\ \bibnamefont
  {Marston}}\ and\ \bibinfo {author} {\bibfnamefont {L.}~\bibnamefont
  {Zhang}},\ }\href {https://doi.org/10.1121/1.4982203} {\bibfield  {journal}
  {\bibinfo  {journal} {The Journal of the Acoustical Society of America}\
  }\textbf {\bibinfo {volume} {141}},\ \bibinfo {pages} {3042} (\bibinfo {year}
  {2017})}\BibitemShut {NoStop}%
\bibitem [{\citenamefont {Lima}\ \emph {et~al.}(2020)\citenamefont {Lima},
  \citenamefont {Le{\~a}o-Neto}, \citenamefont {Marques}, \citenamefont
  {Silva}, \citenamefont {Lopes},\ and\ \citenamefont
  {Silva}}]{lima2020_spheroid}%
  \BibitemOpen
  \bibfield  {author} {\bibinfo {author} {\bibfnamefont {E.~B.}\ \bibnamefont
  {Lima}}, \bibinfo {author} {\bibfnamefont {J.~P.}\ \bibnamefont
  {Le{\~a}o-Neto}}, \bibinfo {author} {\bibfnamefont {A.~S.}\ \bibnamefont
  {Marques}}, \bibinfo {author} {\bibfnamefont {G.~C.}\ \bibnamefont {Silva}},
  \bibinfo {author} {\bibfnamefont {J.~H.}\ \bibnamefont {Lopes}},\ and\
  \bibinfo {author} {\bibfnamefont {G.~T.}\ \bibnamefont {Silva}},\ }\href
  {https://doi.org/10.1103/PhysRevApplied.13.064048} {\bibfield  {journal}
  {\bibinfo  {journal} {Physical Review Applied}\ }\textbf {\bibinfo {volume}
  {13}},\ \bibinfo {pages} {064048} (\bibinfo {year} {2020})}\BibitemShut
  {NoStop}%
\bibitem [{\citenamefont {Jerome}\ \emph {et~al.}(2021)\citenamefont {Jerome},
  \citenamefont {Ilinskii}, \citenamefont {Zabolotskaya},\ and\ \citenamefont
  {Hamilton}}]{jerome2021_spheroid}%
  \BibitemOpen
  \bibfield  {author} {\bibinfo {author} {\bibfnamefont {T.~S.}\ \bibnamefont
  {Jerome}}, \bibinfo {author} {\bibfnamefont {Y.~A.}\ \bibnamefont
  {Ilinskii}}, \bibinfo {author} {\bibfnamefont {E.~A.}\ \bibnamefont
  {Zabolotskaya}},\ and\ \bibinfo {author} {\bibfnamefont {M.~F.}\ \bibnamefont
  {Hamilton}},\ }\href {https://doi.org/10.1121/10.0003813} {\bibfield
  {journal} {\bibinfo  {journal} {The Journal of the Acoustical Society of
  America}\ }\textbf {\bibinfo {volume} {149}},\ \bibinfo {pages} {2081}
  (\bibinfo {year} {2021})}\BibitemShut {NoStop}%
\bibitem [{\citenamefont {Lopes}\ \emph {et~al.}(2020)\citenamefont {Lopes},
  \citenamefont {Lima}, \citenamefont {Le{\~a}o-Neto},\ and\ \citenamefont
  {Silva}}]{lopes2020_spin}%
  \BibitemOpen
  \bibfield  {author} {\bibinfo {author} {\bibfnamefont {J.~H.}\ \bibnamefont
  {Lopes}}, \bibinfo {author} {\bibfnamefont {E.~B.}\ \bibnamefont {Lima}},
  \bibinfo {author} {\bibfnamefont {J.~P.}\ \bibnamefont {Le{\~a}o-Neto}},\
  and\ \bibinfo {author} {\bibfnamefont {G.~T.}\ \bibnamefont {Silva}},\ }\href
  {https://doi.org/10.1103/PhysRevE.101.043102} {\bibfield  {journal} {\bibinfo
   {journal} {Physical Review E}\ }\textbf {\bibinfo {volume} {101}},\ \bibinfo
  {pages} {043102} (\bibinfo {year} {2020})}\BibitemShut {NoStop}%
\bibitem [{\citenamefont {Lima}\ and\ \citenamefont
  {Glauber}(2021)}]{Silva_2021_RBC}%
  \BibitemOpen
  \bibfield  {author} {\bibinfo {author} {\bibfnamefont {E.~B.}\ \bibnamefont
  {Lima}}\ and\ \bibinfo {author} {\bibfnamefont {T.~S.}\ \bibnamefont
  {Glauber}},\ }\href {https://doi.org/10.1121/10.0005625} {\bibfield
  {journal} {\bibinfo  {journal} {The Journal of the Acoustical Society of
  America}\ }\textbf {\bibinfo {volume} {150}},\ \bibinfo {pages} {376}
  (\bibinfo {year} {2021})}\BibitemShut {NoStop}%
\bibitem [{\citenamefont {Zhang}\ and\ \citenamefont
  {Marston}(2011{\natexlab{a}})}]{zhang2011_bessel_neg}%
  \BibitemOpen
  \bibfield  {author} {\bibinfo {author} {\bibfnamefont {L.}~\bibnamefont
  {Zhang}}\ and\ \bibinfo {author} {\bibfnamefont {P.~L.}\ \bibnamefont
  {Marston}},\ }\href {https://doi.org/10.1103/PhysRevE.84.035601} {\bibfield
  {journal} {\bibinfo  {journal} {Physical Review E}\ }\textbf {\bibinfo
  {volume} {84}},\ \bibinfo {pages} {035601} (\bibinfo {year}
  {2011}{\natexlab{a}})}\BibitemShut {NoStop}%
\bibitem [{\citenamefont {Zhang}\ and\ \citenamefont
  {Marston}(2011{\natexlab{b}})}]{zhang2011_angular}%
  \BibitemOpen
  \bibfield  {author} {\bibinfo {author} {\bibfnamefont {L.}~\bibnamefont
  {Zhang}}\ and\ \bibinfo {author} {\bibfnamefont {P.~L.}\ \bibnamefont
  {Marston}},\ }\href {https://doi.org/10.1103/PhysRevE.84.065601} {\bibfield
  {journal} {\bibinfo  {journal} {Physical Review E}\ }\textbf {\bibinfo
  {volume} {84}},\ \bibinfo {pages} {065601} (\bibinfo {year}
  {2011}{\natexlab{b}})}\BibitemShut {NoStop}%
\bibitem [{\citenamefont {Zhang}(2018)}]{zhang2018_reversals}%
  \BibitemOpen
  \bibfield  {author} {\bibinfo {author} {\bibfnamefont {L.}~\bibnamefont
  {Zhang}},\ }\href {https://doi.org/10.1103/PhysRevApplied.10.034039}
  {\bibfield  {journal} {\bibinfo  {journal} {Physical Review Applied}\
  }\textbf {\bibinfo {volume} {10}},\ \bibinfo {pages} {034039} (\bibinfo
  {year} {2018})}\BibitemShut {NoStop}%
\bibitem [{\citenamefont {Fan}\ and\ \citenamefont
  {Zhang}(2019)}]{zhang_2019_bessel_trap}%
  \BibitemOpen
  \bibfield  {author} {\bibinfo {author} {\bibfnamefont {X.-D.}\ \bibnamefont
  {Fan}}\ and\ \bibinfo {author} {\bibfnamefont {L.}~\bibnamefont {Zhang}},\
  }\href {https://doi.org/10.1103/PhysRevApplied.11.014055} {\bibfield
  {journal} {\bibinfo  {journal} {Physical Review Applied}\ }\textbf {\bibinfo
  {volume} {11}},\ \bibinfo {pages} {014055} (\bibinfo {year}
  {2019})}\BibitemShut {NoStop}%
\bibitem [{\citenamefont {Fan}\ and\ \citenamefont
  {Zhang}(2021)}]{zhang2021_phase_bessel}%
  \BibitemOpen
  \bibfield  {author} {\bibinfo {author} {\bibfnamefont {X.-D.}\ \bibnamefont
  {Fan}}\ and\ \bibinfo {author} {\bibfnamefont {L.}~\bibnamefont {Zhang}},\
  }\href {https://doi.org/10.1121/10.0005491} {\bibfield  {journal} {\bibinfo
  {journal} {The Journal of the Acoustical Society of America}\ }\textbf
  {\bibinfo {volume} {150}},\ \bibinfo {pages} {102} (\bibinfo {year}
  {2021})}\BibitemShut {NoStop}%
\bibitem [{\citenamefont {Sepehrirahnama}\ \emph {et~al.}(2021)\citenamefont
  {Sepehrirahnama}, \citenamefont {Oberst}, \citenamefont {Chiang},\ and\
  \citenamefont {Powell}}]{shahrokh2021_WC_ARF}%
  \BibitemOpen
  \bibfield  {author} {\bibinfo {author} {\bibfnamefont {S.}~\bibnamefont
  {Sepehrirahnama}}, \bibinfo {author} {\bibfnamefont {S.}~\bibnamefont
  {Oberst}}, \bibinfo {author} {\bibfnamefont {Y.}~\bibnamefont {Chiang}},\
  and\ \bibinfo {author} {\bibfnamefont {D.}~\bibnamefont {Powell}},\ }\href
  {https://doi.org/10.1103/PhysRevE.104.065003} {\bibfield  {journal} {\bibinfo
   {journal} {Physical Review E}\ }\textbf {\bibinfo {volume} {104}},\ \bibinfo
  {pages} {065003} (\bibinfo {year} {2021})}\BibitemShut {NoStop}%
\bibitem [{\citenamefont {Marston}\ \emph {et~al.}(2006)\citenamefont
  {Marston}, \citenamefont {Wei},\ and\ \citenamefont {Thiessen}}]{Marston}%
  \BibitemOpen
  \bibfield  {author} {\bibinfo {author} {\bibfnamefont {P.~L.}\ \bibnamefont
  {Marston}}, \bibinfo {author} {\bibfnamefont {W.}~\bibnamefont {Wei}},\ and\
  \bibinfo {author} {\bibfnamefont {D.~B.}\ \bibnamefont {Thiessen}},\ }in\
  \href@noop {} {\emph {\bibinfo {booktitle} {AIP Conference Proceedings}}},\
  Vol.\ \bibinfo {volume} {838}\ (\bibinfo {organization} {AIP},\ \bibinfo
  {year} {2006})\ pp.\ \bibinfo {pages} {495--499}\BibitemShut {NoStop}%
\bibitem [{\citenamefont {Fan}\ \emph {et~al.}(2008)\citenamefont {Fan},
  \citenamefont {Mei}, \citenamefont {Yang},\ and\ \citenamefont
  {Chen}}]{fan2008torque}%
  \BibitemOpen
  \bibfield  {author} {\bibinfo {author} {\bibfnamefont {Z.}~\bibnamefont
  {Fan}}, \bibinfo {author} {\bibfnamefont {D.}~\bibnamefont {Mei}}, \bibinfo
  {author} {\bibfnamefont {K.}~\bibnamefont {Yang}},\ and\ \bibinfo {author}
  {\bibfnamefont {Z.}~\bibnamefont {Chen}},\ }\href@noop {} {\bibfield
  {journal} {\bibinfo  {journal} {The Journal of the Acoustical Society of
  America}\ }\textbf {\bibinfo {volume} {124}},\ \bibinfo {pages} {2727}
  (\bibinfo {year} {2008})}\BibitemShut {NoStop}%
\bibitem [{\citenamefont {Sepehrirahnama}\ and\ \citenamefont
  {Lim}(2020{\natexlab{b}})}]{Sh2020_agglo}%
  \BibitemOpen
  \bibfield  {author} {\bibinfo {author} {\bibfnamefont {S.}~\bibnamefont
  {Sepehrirahnama}}\ and\ \bibinfo {author} {\bibfnamefont {K.-M.}\
  \bibnamefont {Lim}},\ }\href@noop {} {\bibfield  {journal} {\bibinfo
  {journal} {Microfluidics and Nanofluidics}\ }\textbf {\bibinfo {volume}
  {24}},\ \bibinfo {pages} {1} (\bibinfo {year}
  {2020}{\natexlab{b}})}\BibitemShut {NoStop}%
\bibitem [{\citenamefont {Mitri}(2009)}]{mitri2009Bessel}%
  \BibitemOpen
  \bibfield  {author} {\bibinfo {author} {\bibfnamefont {F.}~\bibnamefont
  {Mitri}},\ }\href@noop {} {\bibfield  {journal} {\bibinfo  {journal}
  {Ultrasonics}\ }\textbf {\bibinfo {volume} {49}},\ \bibinfo {pages} {794}
  (\bibinfo {year} {2009})}\BibitemShut {NoStop}%
\bibitem [{\citenamefont {Liu}\ and\ \citenamefont
  {Lim}(2011)}]{liu2011_freq_switch}%
  \BibitemOpen
  \bibfield  {author} {\bibinfo {author} {\bibfnamefont {Y.}~\bibnamefont
  {Liu}}\ and\ \bibinfo {author} {\bibfnamefont {K.-M.}\ \bibnamefont {Lim}},\
  }\href@noop {} {\bibfield  {journal} {\bibinfo  {journal} {Lab on a Chip}\
  }\textbf {\bibinfo {volume} {11}},\ \bibinfo {pages} {3167} (\bibinfo {year}
  {2011})}\BibitemShut {NoStop}%
\end{thebibliography}%


\end{document}